\documentclass[aps,prb,groupedaddress,amsmath,amssymb,amsfonts,showpacs,twocolumn]{revtex4}
\usepackage{bm}
\usepackage{graphicx}
\usepackage{color}
\usepackage{psfig}

\begin{document}


\title{Full Counting Statistics of a charge shuttle}
\author{F. Pistolesi}
\affiliation{
Laboratoire de Physique et Mod\'elisation des Milieux Condens\'es,\\
CNRS-UJF B.P. 166, F-38042 Grenoble, France
}

\affiliation{
Argonne National
Laboratory, 9700 S.Cass Ave, Argonne, IL 60439, USA
}

\date{\today}

\newcommand{\qav}[1]{\left\langle #1 \right\rangle}
\newcommand{\av}[1]{\overline{#1}}
\newcommand{\myT}{\Gamma}
\newcommand{\rem}[1]{}
\newcommand{\FORSE}[1]{{\bf Peut-etre #1} }
\newcommand{\refe}[1]{(\ref{#1})}
\newcommand{\refE}[1]{Eq.~(\ref{#1})}
\newcommand{\beq}{\begin{equation}}
\newcommand{\eeq}{\end{equation}}
\newcommand{\cg}{\check g}
\newcommand{\inc}{{\rm inc}}
\newcommand{\beqa}{\begin{eqnarray}}
\newcommand{\eeqa}{\end{eqnarray}}

\begin{abstract}
We study the charge transfer in a small grain oscillating between
two leads.
Coulomb blockade restricts the charge fluctuations in such
a way that only zero or one additional electrons can
sit on the grain.
The system thus acts as a charge shuttle.
We obtain the full counting statistics of charge transfer and
discuss its behavior.
For large oscillation amplitude the probability of
transferring $\tilde n$ electrons per cycle is strongly
peaked around one.
The peak is asymmetric since its form is controlled by different
parameters for $\tilde n>1$ and $\tilde n < 1$.
Under certain conditions the systems behaves as if
the effective charge is 1/2 of the elementary one.
Knowledge of the counting statistics  gives a
new insight on the mechanism of charge transfer.
\end{abstract}



\maketitle

\section{Introduction}
\label{sec1}

\begin{figure}
\centerline{\psfig{file=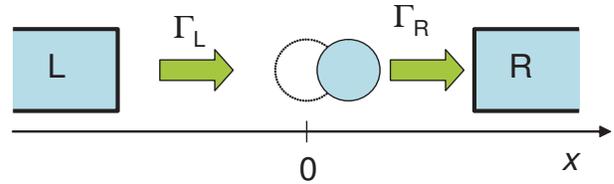,width=8cm}}
\caption{Simple schematization of the system. A small grain oscillating
in an harmonic potential between two leads kept at a constant
voltage bias.}
\label{Fig1}
\end{figure}

A few years ago Gorelik {\em et al.} showed that a small conducting
grain can undergo a mechanical instability if it is trapped in a
harmonic potential between two leads kept at a constant voltage
bias (see Fig. \ref{Fig1}).\cite{Gorelik}
When the central grain is charged, the electrostatic force induced by
the leads pushes it towards one of the electrodes, increasing the
probability that the excess charge be discharged.
Since the resistance depends exponentially on the distance, even
small oscillations can largely amplify the probability of transmission.
Excitation of this nanomechanical system at one of its resonating frequencies
can be generated by the stochastic tunnelling of electrons from the leads.
Since the charge state of the grain is correlated with its
position, under certain conditions, the energy accumulated in the
mechanical systems increases indefinitely, leading to an
instability.
The energy pumped depends on the oscillation amplitude
up to a maximum value determined by the number of charges that
can accumulate in the grain at each cycle.
After that point there is no additional gain in increasing the amplitude
and the grain stabilizes at a fixed oscillation amplitude
for which the energy pumped exactly balances the energy dissipated.
This scenario has been investigated both in the incoherent\cite{Gorelik,Nishi,Boese} and
in the quantum case.\cite{Fedorets,Armour,McCarthy,Novotni,Nord,NewSweden}.
There are strong indications that Parks {\em et al.} have
observed this phenomenon in C$_{60}$ molecules oscillating between
two leads.\cite{Park}

An other possibility to drive the oscillations is to
use an external alternate electric field acting on a
a cantilever.
The amplitude can thus be tuned independently of the
source/drain voltage bias, at least in principle.
This case has been experimentally realized by Erbe {\em et al.}
who observed a current of 0.11 electrons per cycle at low
temperature induced by the oscillation of the central
grain.
When the leads and the grain are superconducting the
existence of phase coherent transport as been
proposed.\cite{GorelikNature,Isacsson,Romito}

Many papers studied theoretically the conditions for the
realization of the instability or considered the dependence of the
current on the external parameters.
Only few papers investigated instead current fluctuations.
Weiss and Zwerger\cite{Weiss} calculated the average number
of electrons transmitted and its fluctuation
during {\em a single} cycle of a shuttle driven at a given
frequency and amplitude.
This quantity differs from the noise actually
measured since typical measurement times are
much longer than one period of oscillation.
Correlations of charge fluctuations
on different cycles are then important, as we
discuss in the following.
Other authors considered the finite frequency noise in superconducting
shuttles\cite{Romito} or the telegraph noise induced by the switching
between two mechanical modes in a two-oscillating-grains
device.\cite{NichiTele}
A related problem is the fluctuation of the acoustoelectric current
carried by surface acoustic waves propagating along a ballistic
quantum channel.\cite{Galperin}

The full counting statistics (FCS) of charge transfer in a shuttle
has not been considered so far.
Recently, powerful techniques have been developed to calculate the
probability than $n$ electrons are transferred during a measurement time
$t_o$ in electronic devices.\cite{Levitov,NazarovMicro,Bagrets,Belzig}
The FCS contains much more information on the dynamics of the
charge transfer than the current or the noise alone.
This will be particularly clear in this problem since few electrons
are involved in the tunnelling, and the probability
that $\tilde n$ electrons per cycle are transmitted is actually
a fundamental quantity.
For a well developed shuttling regime
the noise to current ratio is expected to be
small, since the number of electrons shuttled at every cycle
is determined by the Coulomb blockade conditions and thus
it does not fluctuate as it happens in a purely stochastic tunnelling.
This implies that the probability distribution has a small
width, but its actual shape still depends on
the physical parameters of the junction, like voltage bias,
tunnelling probability, or oscillation amplitude of the shuttle.
The importance of studying theoretically the FCS is thus twofold:
first it is, at least in principle, a measurable quantity
and secondly, its knowledge allows to infer detailed
information on the mechanism of charge transfer.

The paper is organized as follows.
In Sec. \ref{sec2} the technique for the calculation is developed and
the equations for the numerical approach are obtained.
In Sec. \ref{sec3} the FCS is calculated analytically for
small and large oscillations.
In Sec. \ref{sec4} the general (numerical) results are discussed and
compared with the analytical ones.
Section \ref{sec5} gives our conclusions.

\section{Full counting statistics for an oscillating grain}
\label{sec2}

Our aim is to calculate the FCS of charge transfer in an oscillating
grain between two leads.
We assume that the charge transfer is described by
the ``orthodox theory'' of Coulomb blockade.\cite{CBpioneer,CB}
In this regime the tunnelling is incoherent and the dynamics
is governed by a standard master equation.
Within these assumptions Bagrets and Nazarov have developed an
elegant and efficient technique to derive
the FCS for the static case.\cite{Bagrets}
We will use their method generalized to a moving grain.

Since we are interested to a single grain structure,
the state of the system is completely determined by the
probability $p_k$ of having $k$ additional electrons in the
island.
For simplicity we consider the case where the voltage biases
guarantee that only the two states, $k=0,1$,
are available, and that only two events are possible:
either one electron jumps on the island from the left lead
[with transition rate $\Gamma_L(t)$] or one electron on the
island (if present) jumps to the right lead
[with transition rate $\Gamma_R(t)$].
Within these assumptions the time evolution of the probability is given
by a master equation.
We write it following the notation of Ref. \onlinecite{Bagrets}:
\beq
{\partial \over \partial t} |p (t) \rangle
=
-\hat L(t)  |p(t)\rangle
\eeq
where the (classical) probability is represented by a state in a vector space:
$|p\rangle = \{ p_0, p_{1} \}$, and $\hat L$ is the matrix
\beq
\hat L(t) =
\left( \begin{array}{cc} \Gamma_L(t) & -\Gamma_R(t)
\\
-\Gamma_L(t) & \Gamma_R(t)
\end{array} \right)
\label{firstL}
\, .
\eeq

In Ref. \onlinecite{Bagrets} it is shown how the FCS can be obtained
by calculating the time evolution of the probability with a modified
operator $\hat L(t)$.
The central quantity is $P_{t_o}(n)$, the probability that $n$ electrons
have been transmitted during a measurement time $t_o$.
This quantity is independent from the initial condition in the limit of
large $t_o$.
From the technical point of view it is easier to calculate the
generating function $S_{t_o}(\chi)$:
\beq
    e^{-S_{t_o}(\chi)} = \sum_{n=0}^\infty P_{t_o}(n) e^{i n \chi}
    \,.
    \label{defS}
\eeq
From $S$ one can easily obtain all cumulants:
$\av{n}=\partial(-S)/\partial(i\chi)|_{\chi=0}$,
$\av{(n-\av{n})^2} = \partial^2 (-S)/\partial(i\chi)^2|_{\chi=0}   $, etc.
Let us count electrons crossing, for instance, the left junction.
According to the prescriptions of Ref. \onlinecite{Bagrets}
the modified operator $\hat L_\chi(t)$ is obtained from \refE{firstL}
by multiplying the lower off-diagonal matrix element by
the factor $e^{i \chi}$.
This factor keeps track of the electrons that
cross the left junction during the time evolution.
The generating function $S_{t_o}(\chi)$ is then simply given by the
formal integration of the modified time-evolution equation:
\beq
e^{-S_{t_o}(\chi)}
=
\langle q |
{\rm  Texp}\left\{- \int_0^t \hat L_\chi(t') dt' \right\}
|p(0)\rangle
\label{gener}
\,,
\eeq
where $|p(0)\rangle$ is the probability at time $t=0$,
$|q\rangle\equiv \{1,1 \}$, and Texp is the time ordered
exponential.
The derivation of this equality was done in
Ref. \onlinecite{Bagrets} for the static case where $\Gamma$'s
do not depend on time.
Following the steps of their proof it is not difficult to verify
that \refE{gener} holds also in the dynamical case of interest here.
The main difference is that for the static case (and the zero
frequency noise) one can restrict to the study of the eigenvalues
of $\hat L$, since the time ordering becomes immaterial when $\hat L$
does not depend on time.
In our case instead time-ordered exponential must be evaluated
explicitly.

\subsection{Specific expressions for the shuttle}

Let us now consider explicitly the time dependence of
the tunnelling rate.
With good accuracy one can assume that the dependence of
$\Gamma_{L/R}$ on the position of the grain is exponential
\beq
    \Gamma_{L/R}
    =
    \Gamma^0 \exp\{ \mp x/\lambda \}
\eeq
(we assume $\Gamma_L=\Gamma_R=\Gamma^0$ for $x=0$).
Here $x$ is the shift of the grain from the equilibrium position and
$\lambda$ is the tunnelling length (see also Fig. \ref{Fig1}).
We will consider the case of  sinusoidal oscillations of the grain.
This can be driven by an external device like in the experiment of
Ref. \onlinecite{Erbe}, or it can be induced by the voltage bias
between the left and right leads.\cite{Gorelik,Park}
In both cases
\beq
    \Gamma_{L/R}(t) = \Gamma^0 \exp\{ \mp a \sin(\omega t) \} \, ,
\eeq
where $a=x_{max}/\lambda$ is the dimensionless ratio of the oscillation amplitude to
the tunnelling length  and $\omega$ is the
frequency of oscillation.
It is also convenient to rescale the time by $\omega^{-1}$ and define
$\phi = \omega t$.
With this substitution the problem is fully characterized by the two
dimensionless parameters: $a$ and $\Gamma=\Gamma^0/\omega$.
From the physical point of view, $\Gamma$ gives the probability that
an electron in the static junction with $x=0$ can tunnel on or off the
grain in the time $1/\omega$.
We will see that the dependence on $a$ of the FCS will be qualitatively
different if $\Gamma$ is smaller or larger than one.

The interesting physical quantity is the FCS for a long measurement
time $t_o$.
We choose $t_o$ to be a multiple of the period: $t_o=2\pi N/\omega$, with $N$ integer.
The FCS of charge transfer during $N$ periods is then given by
\beq
    e^{-S_N(\chi)}
    =
    \langle q | {\hat A}^N |p(0)\rangle
\eeq
where
\beq
    \hat A = {\rm  Texp}\left\{- \int_0^{2 \pi } \hat L_\chi(\phi') d\phi' \right\}
\eeq
and
\beq
    \hat L_\chi(\phi) =
    \left(
    \begin{array}{cc}
        \Gamma e^{-a \sin \phi}  & -\Gamma e^{a \sin \phi}
        \\
        -\Gamma e^{-a \sin \phi} e^{i \chi}  & \Gamma e^{a \sin \phi}
    \end{array}
    \right)
    \label{Lphi}
    \, .
\eeq
In the case of interest of large $N$ the FCS is given by the
eigenvalue $\lambda_M(\chi)$ of $\hat A$ that has the maximum absolute
value: $ S_N(\chi) = -N \ln[\lambda_M(\chi)] $.
From $S_N(\chi)$ one can calculate directly the particle current and
noise reduced to a period:
$I = \av{\tilde n} = \av{n}/N$ and
$P = 2 \av{(n-\av{n})^2 }/N$.

To obtain the probability of having transferred $n$ electrons during
$N$ periods it suffices to invert \refE{defS}:
\beq
    P_N(n)
        =
        \int_{-\pi}^{+\pi} {d \chi \over 2 \pi} \,
        e^{-S_N(\chi)-i \chi n}
        \, .
\eeq
For large $N$ the saddle point approximation gives a very accurate
estimate of this integral:
\beq
    \ln [P_N(n)]/N = \ln [\lambda_M(\chi_0)] - i \chi_0 \tilde n
    \label{saddleP}
\eeq
where $\tilde n=n/N$ is the number of electrons transferred per cycle and
$\chi_0$ satisfies the equation:
\beq
    {1\over \lambda_M(\chi_0)}
    \left.{d \lambda_M \over d \chi}\right|_{\chi=\chi_0} = i \tilde n.
    \label{saddle}
\eeq
We find that \refE{saddle} is solved by $\chi_0$ pure imaginary.

The problem is now reduced to the evaluation of the time-ordered product
that enters the definition of $\hat A$.
This can be done numerically by integrating the system of
differential equations
\beq
  {\partial \over \partial \phi}  |p(\phi) \rangle
  =
  -\hat L_\chi(\phi) | p(\phi) \rangle
    \label{difeq}
\eeq
with the two initial conditions $|p^{(1)}(\phi=0)\rangle =\{1,0\}$
and $|p^{(2)}(\phi=0)\rangle =\{0,1\}$.
One can readily verify that the matrix with columns
$|p^{(1)}(\phi=2\pi)\rangle$ and $|p^{(2)}(\phi=2 \pi)\rangle$
coincides with $\hat A$.
In the case of interest the numerical task is not hard,
nevertheless discussion of tractable analytical limits greatly
enhances the understanding of the results.
We thus discuss in the next section the small and large amplitude limits
before presenting the numerical results for the general case in
Sec. \ref{sec4}.

\section{Analytical limits for small and large amplitude}
\label{sec3}

\subsection{Static case and fractional charge}

For $a=0$ we have a standard static single electron transistor.
Since $\hat L_\chi$ does not depend on $\phi$,
the time-ordered exponential becomes a simple exponential
\beq
    \hat A = e^{-2 \pi \hat L_\chi}
\eeq
and the generating function for large $N$ can be obtained by
diagonalization of the $\hat L_\chi$ matrix.
The smallest eigenvalue in modulus gives $S_N(\chi)$ at the leading order.
The generating function reads
\beq
   -S_N(\chi)/N =  2 \pi \Gamma \left( e^{i \chi /2}-1 \right)
   \,,
\label{static}
\eeq
in agreement with the result obtained with a different technique
in Ref.~\onlinecite{deJong}.
The current and the noise are thus: $I=P=\pi \Gamma$ with
a Fano factor $F=P/2I$ equal to 1/2.

\begin{figure}
\centerline{\psfig{file=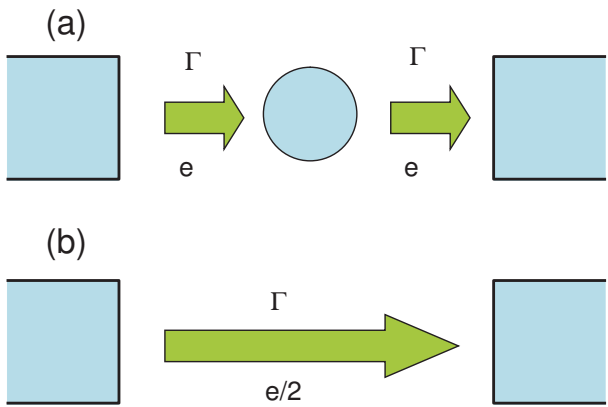,width=8cm}}
\caption{The two equivalent systems for the statistics of
charge transferred in the long measurement time limit:
 (a) single electrons transistor with equal probability $\Gamma$
of hopping from the left lead to the grain and from the grain to the right lead
(b) a fictitious system of charges 1/2 that tunnel with probability $\Gamma$.}
\label{Fig2}
\end{figure}

Even if \refE{static} has been derived before its
meaning has not been fully discussed and it is
worth a short digression.
Tunnelling through a single barrier is a Poissonian process.
The generating function and the probability in this case is
\beq
    -S_{t_o}(\chi) = \av{n} \left( e^{i \chi}-1 \right)
    \quad
    {\rm and}
    \quad
    P_{t_o}(n) = e^{- \av{n}} {\av{n}^n \over n!}
    \,,
\eeq
with $\av{n}$ the average number of charges transmitted during the time $t_o$.
A general feature of the generating function is the $2\pi$ periodicity.
It is a manifestation of the discrete nature of the charge and follows
directly from the definition \refe{defS}.
It is thus surprising that the generating function \refe{static} is
periodic of $4\pi$ as if the elementary charge was not one, but
1/2.
This happens only when $\Gamma_L=\Gamma_R$ and for long measurement times
($N\rightarrow \infty$), in all other cases $S(\chi)$ is periodic of
$2\pi$.
For instance if $\Gamma_L\neq \Gamma_R$, even in the large $N$ limit we have
the following $2\pi$-periodic generating function
\beq
    {S_N(\chi) \over  N (2\pi/\omega)}
    = {\Gamma_L+\Gamma_R \over 2}
    -\sqrt{ {(\Gamma_L-\Gamma_R)^2\over 4} + \Gamma_L \Gamma_R e^{i \chi}}
    \,.
\eeq
The result \refe{static} for $\Gamma_L=\Gamma_R$ indicates that the
charge transfer in our system (Fig. \ref{Fig2} a) is equivalent to that
happening in a tunnel junction with charges 1/2 emitted with probability $\Gamma_0$
(Fig. \ref{Fig2} b).

This can be understood with simple arguments.
Let us consider a sequence of events in our system.
These can be of two types, either tunnelling from
the left lead to the grain (type L),
or tunnelling from the grain to the right lead (type R).
Since the rates for both events are the same
($\Gamma^0$), the system has the same probability
per unit time to switch to the other state.
The statistics of the switching events
(i.e. that either R or L occurs, without
specifying which one) follows thus a Poissonian distribution,
since all events are independent (rates do not depend on the
initial states).

To obtain the statistics of charge transfer from
the statistics of switching events it is enough
to remember that every two switching events one
charge is transmitted.
We can thus associate the transmission of
a fictitious 1/2 charge at each switching event.
It is clear that the counting statistics of
the fictitious charge coincides with that
of the true charge, apart from a possible
charge 1/2 mismatch that is irrelevant for long
measurement times ($t_o \Gamma^0 \rightarrow \infty$).
Thus we proved that the counting statistics
of our system coincides with that
predicted for a tunnelling junction of charges 1/2
transmitted with rate $\Gamma^0$:
simple classical correlations can induce
current fluctuations typical for fractional charges!
A similar "fractional" behavior was also found and discussed
by Andreev and Mishchenko for charge pump in the Coulomb
blockade regime.\cite{Andreev}

One should keep in mind that for any finite measurement
time the periodicity of the generating function
remains $2\pi$, it is only the leading term for $t_o\rightarrow \infty$
that is $4\pi$ periodic.
We thus expect that the Fourier series \ref{defS} is not uniformly convergent and that
for any finite $t_o$ higher moments will definitely depart from the
prediction obtained with \refE{static}.
This can be verified by calculating the first correction
to \refE{static} (we recall that $2\pi N \Gamma = t_o \Gamma_o$):
\beq
    -{S_N\over 2 \pi N \Gamma }
    =
    \left(e^{i \chi/2}-1\right)
    + {1\over 2 \pi N \Gamma} \log(1+\cos \chi/2)
    +\dots
    \,,
    \label{gen2}
\eeq
that holds with accuracy $O(e^{-2\pi N \Gamma})$.
From \refE{gen2} we derive all moments $M_q \equiv {\partial^q (-S) / \partial (i\chi)^q}$.
For $q$ even the result is
\beq
    M_q =
    {2 \pi N \Gamma \over 2^q}\left[ 1 + {(-1)^{q/2+1} \over 2\pi N \Gamma }
    {4 \zeta(q)\over \pi^q}
    (1-2^{-q})(q-1)! \right]
    \, ,
    \label{moments}
\eeq
while for $q$ odd the correction in \refE{gen2} gives
a vanishing contribution.
For large $q$ and fixed $N$ the second term in \refE{moments}
is approximatively $ 4(q/e\pi)^q/(2\pi N\Gamma$).
This means that for $q$ large enough
$M_q$ will depart from the prediction of \refE{static}.
This is the way in which the system may reveal the
true nature of the elementary charge.
Either by a short time measurement of the first moments, or by the
long time measurement of higher moments.

To explain while for asymmetric tunnelling rates
the periodicity is $2 \pi$ even for large $t_o$ it is
enough to notice that the above discussed mapping cannot be realized when
$\Gamma_L \neq \Gamma_R$.
It is worth mentioning that for not too large asymmetries
the fractional charge remains measurable, as it is clear
from the dependence of the Fano factor:\cite{Chen, Korotkov,deJong}
\beq
    F = { \Gamma_L^2+\Gamma_R^2 \over (\Gamma_L+\Gamma_R)^2 }
    \label{FanoStatic}
    \,.
\eeq
But again one expects that departure from the prediction of \refE{static}
will increase with the order $q$ of the moment, and for any small
asymmetry it will become large for $q$ large enough.
Generating functions with periodicity induced by smaller fractions of the
elementary charge can be obtained with several islands.

\subsection{Large oscillation amplitude}

Let us now consider the opposite limit of large oscillation amplitude
of the shuttle.
In this limit, since for most of the time the ratio
$\Gamma_L/\Gamma_R$ is either very large or very small,
we can assume that ({\em i})
for
$0 < \phi < \pi$ the quantity $\Gamma_L$ vanishes identically
and
({\em ii})
 for
$\pi < \phi < 2 \pi$ the opposite holds: $\Gamma_R=0$.
The approximation becomes exact for
$\Gamma \ll 1$, since in that case electrons
can tunnel only when the shuttle is near one of the
two leads.

Within this approximation $\hat A$ can be obtained
analytically.
As a matter of facts in region ({\em i}) $p_0(\phi)+p_1(\phi)$ is
conserved, since the matrix element $\hat L$ that multiplies
$e^{i \chi}$ vanishes.
(We recall that $p_k(\phi)$ is the probability
that $k$ electrons are present in the  grain at time $\phi$.)
The introduction of the counting field normally breaks the conservation
of the probability $p_0(\phi)+p_1(\phi)=const$.
Using this conservation and integrating the remaining differential equation,
we obtain for region ({\em i})
\beq
\left\{
\begin{array}{rcl}
    p_0(\pi) &=& p_0(0)+(1-\alpha)p_1(0)
    \\
    p_1(\pi) &=& \alpha p_1(0)
\end{array}
\right.
\eeq
where $1-\alpha$ is the probability of transferring one electron
during half cycle given with
\beq
    \alpha = \exp\left\{-\Gamma \int_0^\pi e^{ a \sin \phi} d\phi \right\}
    \,.
\eeq
In region ({\em ii}) $p_0(\phi)+e^{-i \chi} p_1(\phi)$ is conserved and we
find
\beq
\left\{
\begin{array}{rcl}
    p_0(2\pi) &=& \alpha p_0(\pi)
    \\
    p_1(2 \pi) &=& p_1(\pi)+(1-\alpha)^p_0(\pi) e^{i \chi}
    \quad .
    \end{array}
\right.
\eeq
By composing the evolution in ({\em i}) and ({\em ii}) we find
\beq
    \hat A
    =
\left(
\begin{array}{cc}
    \alpha & \alpha(1-\alpha) \\
    (1-\alpha)e^{i \chi} & \alpha+(1-\alpha)^2 e^{i \chi}
\end{array}
\right)
\label{Amatrix}
\eeq
The generating function
is obtained by diagonalization of \refE{Amatrix}:
\beq
\lambda_M
=
\alpha
+(1-\alpha) {y\over 2} \sqrt{(1-\alpha)^2 y^2 + 4 \alpha}
+(1-\alpha)^2 {y^2\over 2}
\,,
\label{FCSalpha}
\eeq
where we introduced the short hand notation
$y=e^{i \chi/2}$.
Current and noise follows by differentiation:
\beq
    I = {1-\alpha \over 1+ \alpha }\,,
    \qquad
    P = 4\alpha {1-\alpha \over (1+\alpha)^3}
    \,.
\eeq

For $\Gamma\ll 1$ \refE{FCSalpha} is very accurate
and holds for $0\leq \alpha \leq 1$.
It is instructive to study its behavior in the two opposite limits of
$\alpha\ll 1$ and $1-\alpha \ll 1$.

For $\alpha\ll 1$ the probability of transferring the
electron during the half cycle is nearly 1.
Linearizing \refE{FCSalpha} in $\alpha$ we find the
generating function of a binomial distribution:
\beq
    e^{-S_N(\chi)} = [2\alpha+(1-2\alpha) e^{i \chi}]^N
    \,.
    \label{smallal}
\eeq
This means that at each cycle one electron
is transmitted with probability $1-2\alpha$.
The cycles are independent: after $N$ cycles the probability of
having transmitted $n$ electrons is simply given by the binomial
distribution $\left( {N \atop n}\right) (1-2\alpha)^n (2
\alpha)^{N-n}$.
Cycles are independent for $\alpha\rightarrow 0$ since at each cycle
the system is reset to the stationary state within accuracy $\alpha^2$,
regardless of the initial state.
The stationary solution is given by the eigenvector of \refE{Amatrix}
with eigenvalue 1 for $\chi=0$: $|p_{st}\rangle = \{\alpha/(1+\alpha), 1/(1+\alpha)\}$.
Calculating the transmission probability for one electron during one
cycle with initial condition given by $|p_{st}\rangle$ one obtains, with linear
accuracy, the correct result $1-2\alpha$ appearing in \refE{smallal}.

For $\alpha\rightarrow 1$ the probability for one electron to
tunnel during a cycle is very small.
We can thus expand the generating function in the positive quantity
$(1-\alpha) \ll 1$.
This gives the following surprising result:
\beq
    e^{-S_N(\chi)} = \left[\alpha +(1-\alpha) e^{i \chi/2}\right]^N
    \,.
    \label{alpha1}
\eeq
We find again that the periodicity of the generating function
has changed.
\refE{alpha1} describles a system of 1/2 charges that at each
cycle have a probability $1-\alpha$ of being transmitted.
The situation is similar to the static case.
We can again create a mapping on a fictitious system of charges
1/2 and say that every time that one electron succeeds in jumping
on or off the central island, one charge 1/2 is transmitted in
the fictitious system.
This is possible, since it is extremely unlikely that
one electron can perform the full shuttling in one
cycle.
Thus after many cycles ($N \gg 1$) the counting statistics of these
two systems coincide.
The cycles are no more independent like in the case for
$\alpha \ll 1$, but the problem can be mapped onto an
independent tunnelling one.
For $\alpha$ intermediate it is more difficult to give a simple
interpretation of \refE{FCSalpha}, since different cycles
are correlated in a non trivial way.

In Ref. \onlinecite{Weiss} the current and noise within a
similar model have been calculated, but only for a single
cycle using the the stationary solution as initial condition.
This approach clearly neglects correlations among different cycles.
We have seen that this is an excellent approximation for
$\alpha\rightarrow 0$: \refE{smallal} represents $N$ uncorrelated cycles.
But it fails completely in the
opposite limit of $\alpha \rightarrow 1$,
where the main contribution to the current fluctuations
comes from the cycle-cycle correlations.
This can be seen as follows.
Starting from the stationary solution for $\alpha \rightarrow 1$ (i.e. $\{1/2,1/2\}$)
one can calculate the average number of particles transmitted $\av{n}=(1-\alpha)/2$
and its fluctuation $\av{(n-\av{n})^2}=(1-\alpha)/2$ during a {\em single} cycle.
From \refE{alpha1} we see that the average current
over a large number of cycles is correctly reproduced,
but the noise differs by a factor of 2.
This difference increases with higher moments.
Even if the fluctuation during a single cycle is an
interesting physical quantity, the experimentally relevant one
is the long time fluctuations.

\subsubsection*{"Mixing" regions}

We expect that \refE{FCSalpha} describes pretty well the counting
statistics $P_N(n)$ for $n<N$, but it is clear that it fails
completely for $n>N$ for which it gives $P_N(n>N)=0$ identically.
As a matter of fact, the approximation does not take into
account than more than one electron per cycle can be transmitted.
This is an artefact of the assumption that $\Gamma_L$
and $\Gamma_R$ are never non-vanishing at the same time.

\begin{figure}
\centerline{\psfig{file=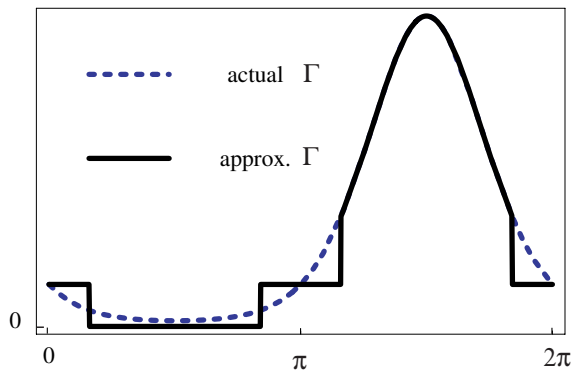,width=7.5cm}}
\caption{Exact and approximate dependence on $\phi$ of $\Gamma_L$
used to obtain analytically the FCS in the large oscillation limit.
$\Gamma_R(\phi)$ has the same form shifted by $\pi$ in the $\phi$ axis.}
\label{Fig3}
\end{figure}

In order to improve the approximation, but keeping the problem
solvable analytically, we need to treat differently the left and right
``mixing'' regions $\phi\approx 0,\ \pi,\ 2 \pi$.
In these regions at lowest order $\Gamma_L \approx \Gamma_R$.
We thus divide the time evolutions in 5 steps.
For $\phi<\phi_0$, $|\phi-\pi|<\phi_0$, and $2\pi-\phi<\phi_0 \sim 1/a \ll 1$
we calculate the evolution with $\Gamma_L=\Gamma_R=\Gamma^0$.
For $\phi_0 < \phi< \pi-\phi_0$ and
$\pi+\phi_0 < \phi< 2\pi-\phi_0$
we use instead the previous approximation for regions
({\em i}) and ({\em ii}).
The approximation is summarized in Fig. \ref{Fig3} where the exact and
the approximate dependence of $\Gamma_L(\phi)$ is shown.

The contribution to $\hat A$ of the three mixing regions depends on
$\phi_0$ and $\Gamma$ only through their product $\Gamma \phi_0 \equiv
\tau$.
The approximation is meaningful only for $\alpha$ small,
otherwise the constant approximation for the probabilities
in the mixing regions would not be accurate.
We thus consider only the small $\alpha$ limit.
Keeping linear terms in $\alpha$ we have
\beqa
\lambda_M(\chi)
    &=&
    {e^{-4\tau}\over 2}
\left[
(1-2\alpha)(y^2-1) + 2 y \sinh(4\tau y)
\right.
    \nonumber \\
&&
\left.
+\left( 1+2\alpha+(1-2 \alpha)y^2\right)\cosh(4 \tau y)
\right]
\label{genAsm}
\eeqa
with the same short hand notation $y=e^{i \chi/2}$.

This expression, through $-S_N(\chi)/N=\ln \lambda_M(\chi)$
gives the FCS for large $a$ in different limits.
Let us begin with the case $\tau \ll 1$.
Expanding \refE{genAsm} up to second order in $\tau$ we obtain
\beq
    e^{-S_N(\chi)}
    =
    \lambda_M(\chi)^N
        =
\left[
    \beta_0
    +   \beta_1 e^{i \chi}
    +   \beta_2 e^{2 i \chi}
\right]^N
    \label{expans}
\eeq
with
\beq
\left\{
\begin{array}{rcl}
\beta_0 &=& 2\alpha(1-4\tau+ 8\tau^2) \\
\beta_1 &=& 1-2\alpha(1-4\tau+4\tau^2)-4\tau^2\\
\beta_2 &=& 4\tau^2 (1-2 \alpha)
 \quad .
\end{array}
\right.
\label{smalltau}
\eeq

The interpretation is again simple, \refE{expans} gives a trinomial
distribution, at each cycle there is a probability $\beta_{\tilde n}$ of
transmitting $\tilde n$ electrons per cycle.
This approximation holds for any $0 < \tilde n < 2$ and $\tau \approx \Gamma/a \ll 1$, $a \gg 1$.
In this case the probability of transmitting more than
two electrons within a cycle is extremely small since
$\beta_{\tilde n>1} \sim \tau^{2\tilde n-2}$.
The importance of the parameter $\tau$ for $\beta_{\tilde n>1}$ proves
that to understand the probability of charge transfer for $n>N$ it is
crucial to correctly treat the mixing regions where $\Gamma_L$ and
$\Gamma_R$ are both non-vanishing.
This means that in this limit the FCS for $\tilde n=n/N > 1$ is
determined mainly by the value of $\tau$, while for $0< \tilde n
<1$ is $\alpha$ that controls the FCS.

Using the saddle point approximation, it is possible to
obtain explicitly the probability for $\tilde n \approx 1$:
\beq
    {\ln\left[ P_N(\tilde n)\right] \over N}
    =
    \left\{
    \begin{array}{ll}
\ln \beta_1 -(1-\tilde n) \ln
\left(
{1-\tilde n \over \beta_0/\beta_1}
\right) &  {\rm for\ }\tilde n <1
    \\
\ln \beta_1 -(\tilde n-1) \ln
\left(
{\tilde n -1 \over \beta_2/\beta_1}
\right) & {\rm for\ } \tilde n  >1 \, .
    \end{array}
    \right.
    \label{prob}
\eeq
The probability has a sharp maximum at $\tilde n=1$, as expected, and
its logarithm decreases approximately linearly on both sides, with
slopes controlled by two different parameters.
For $\tilde n<1$ the slope is approximately given by
$-\ln(2 \alpha)$, while for $\tilde n>1$ it is given
by $\ln(4 \tau^2)$.
Since the parameter $\alpha$ decreases exponentially with $a$, while
$\tau$ is only inversely proportional to $a$, the peak around $\tilde
n=1$ is asymmetric with an excess to the left for moderately large
$a$, and with an excess to the right for larger $a$.

When $\tau\gg 1$ the previous expansion in $\tau$
cannot be used, but in the small region $\tilde n \approx 1$ we can
find analytically the FCS expanding $\lambda_M(\chi)$ in powers of $y$.
In fact, it turns out that the saddle point equation
\refe{saddle} is solved by $\chi=i x$ with $x$ real and large, thus
with $y\ll 1$.
We find that the expansion of $\lambda_M$ contains only even
powers of $y$ and at the fourth order coincide with \refE{expans},
but with the $\beta$ coefficients given by (for large $\tau$)
\beq
\left\{
\begin{array}{rcl}
\beta_0 &=& 2\alpha e^{-4\tau} \\
\beta_1 &=& 4(1+2\alpha) e^{-4 \tau} \tau^2 \\
\beta_2 &=& {16\over 3}(1+2\alpha)  e^{-4 \tau} \tau^4 \quad .
\end{array}
\right.
\label{largetau}
\eeq
For small $y$ it exists a region $\beta_0/\beta_2 < y^2 <
\beta_1/\beta_2$ where the $\beta_1 y^2$ term dominates the other two
terms.
We thus find again the same behavior of \refE{prob} for the
probability, but with the coefficients given by \refE{largetau}.
Note that now $\beta_2>\beta_1$, this means that the probability
of transmitting more than one electron per cycle is always larger
than the probability of transmitting less than one per cycle.
Actually for large $a$ the asymmetry is extreme, the slope for
$\tilde n<1$ is much larger than the slope for $\tilde n>1$ which
is moderately positive.

For $\tilde n \gg 1$ we cannot expand anymore for small $y$.
Since $\alpha$ is not crucial to understand this region we can set
$\alpha=0$ into \refE{genAsm}:
\beq
    \lambda_M(\chi)
    =
    e^{-4\tau}
    \left[ y \cosh(2 \tau y)+ \sinh(2 \tau y) \right)]^2
    \,.
\eeq
For large real $y$ we thus find that the generating function
is that of a static grain active [cfr. \refE{static}] for a
fraction $4\phi_0/(2\pi)$ of the time:
\beq
    -S(\chi)/N = 4 \phi_0 \Gamma\left( e^{i \chi/2} -1 \right)
    \,.
\eeq

\section{General results}
\label{sec4}

The results discussed above can be now compared with the numerical
results valid for arbitrary values of the amplitude $a$.
These are obtained by solving numerically the system of differential
equations \refe{difeq} to calculate $\hat A$.
The matrix is then diagonalized and the maximum eigenvalue in modulus selected.
Current and noise are obtained by numerical differentiation,
while the FCS is obtained by solving numerically \refE{saddle}.

\begin{figure}
\centerline{
\psfig{file=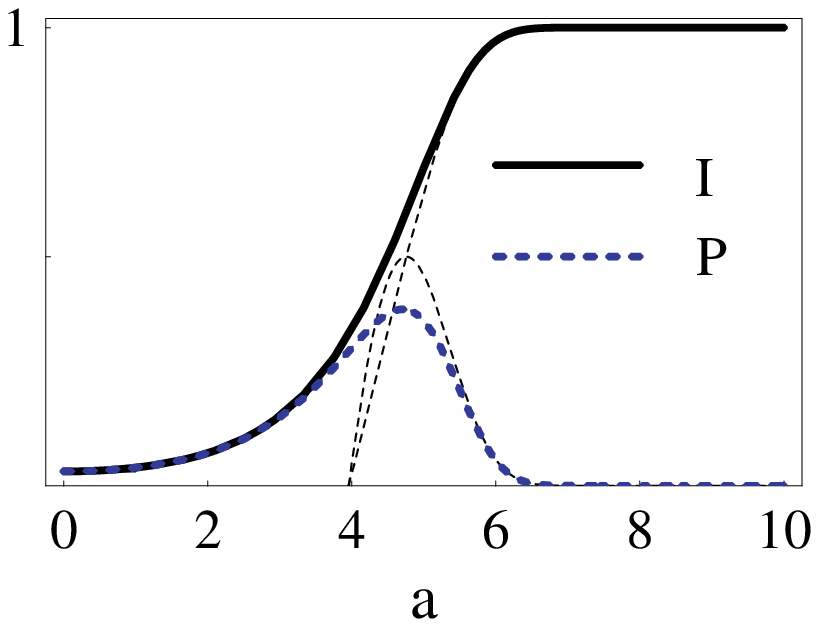,width=4.3cm}
\psfig{file=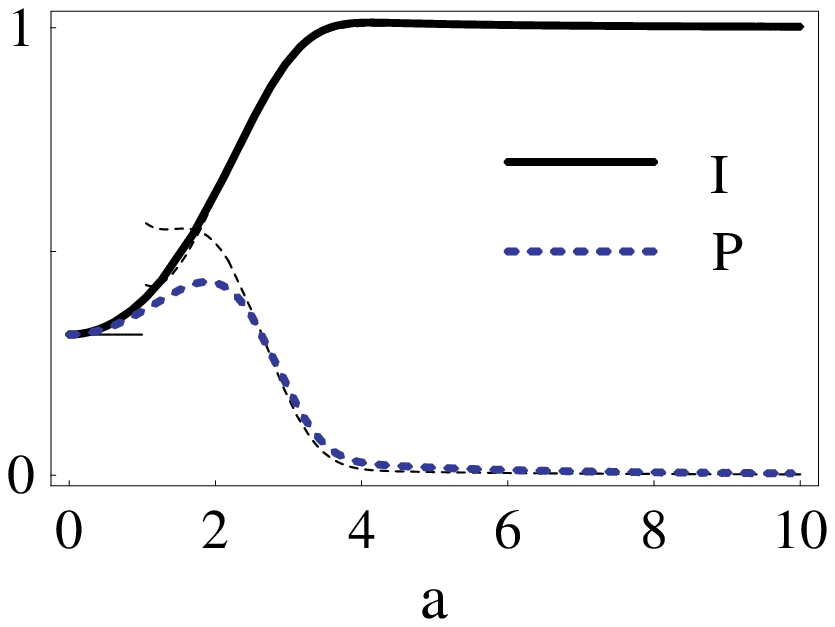,width=4.3cm}
}
\centerline{
\psfig{file=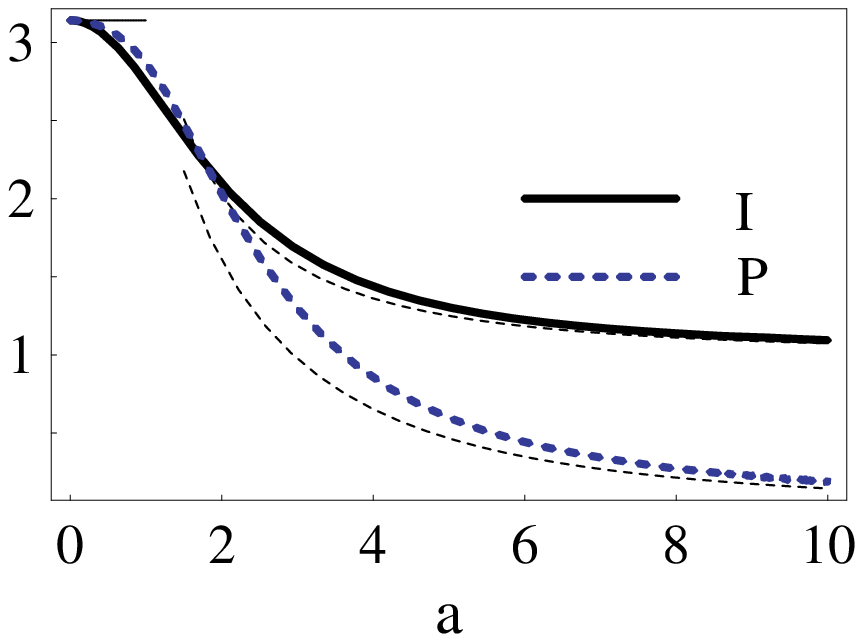,width=4.3cm}
\psfig{file=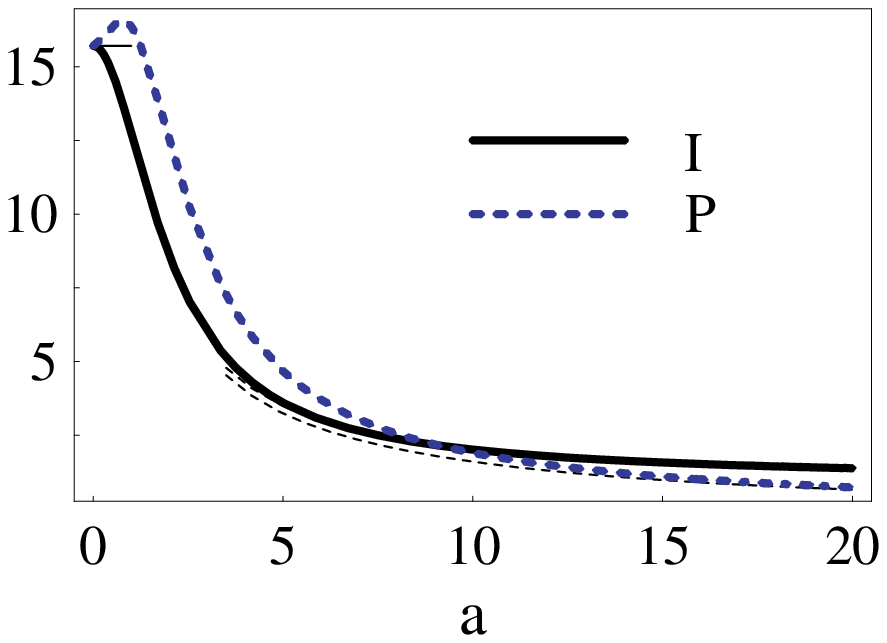,width=4.3cm}
}
\caption{Current (plain line)  and noise (dashed line) as a function of the oscillation
amplitude for different values of $\Gamma$.
From the top-left pannel $\Gamma=$ 0.01, 0.1, 1, 5.
The dashed light lines are obtained with the analytical approximation
\refe{genAsm}.
The short lines at $a=0$ are the static results given by
\refe{static}.
}
\label{FigNoise}
\end{figure}

We begin by discussing current and noise.
Fig. \ref{FigNoise}  shows the average number of electrons and its fluctuation
for different values of $\Gamma$ as a function of the amplitude $a$.
We first notice the qualitative difference between $\Gamma$ smaller or
larger than 1.
In the first case the oscillation of the central grain largely
increases the current, while in the second case it reduces it.
In both cases for large $a$ the current saturates towards one electron
per cycle.
From our previous analysis we know that for large $\Gamma$ the saturation
happens only for very large $a$, when $\tau = 2 \Gamma/a$ becomes small
enough to reduce the contribution of the central region.
(The choice of a factor 2 into the definition of $\tau$ is arbitrary and
it simply improves the accuracy of the analytical approximation.
Any factor of the order of one does not change significantly the
results.)
A striking feature that appears from the plot for the noise is
the enormous reduction of the Fano factor.
The transport becomes deterministic due to
the shuttling, it is very difficult that the grains perform an
oscillation without transmitting one electron.

We believe that measuring noise and current in a device
can give a clean indication if the system is actually shuttling
electrons.
It can discriminate between a simple coupling between the
mechanical and the electronic degrees of freedom of the system
not associated with the shuttling mechanism.

In Fig. \ref{FigNoise} we also plot the comparison with our simple analytical
approximation for large and small $a$.
The agreement is pretty good, indicating that the crucial features
are correctly reproduced by our simple picture of
evolution in five steps.

\begin{figure}
\centerline{\psfig{file=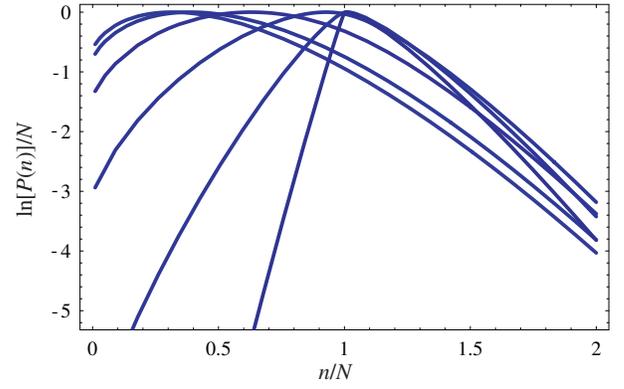,width=8cm}}
\caption{$\ln[P_N(\tilde n)]/N$  as a function of the number of electrons
transferred per cycle $\tilde n$ for $\Gamma=0.1$, and oscillation amplitude
$a=0$, 1, 2, 3, 4, and 5 (from left to right).}
\label{FigFCS1}
\end{figure}

Let us now discuss the counting statistics.
In Fig. \ref{FigFCS1} we show the evolution of $\ln[P_N(\tilde n)]/N$
when  $a$ is increased from 0 to 5 by steps of one unit.
We show the case $\Gamma=0.1$ that is a good representative of
the small $\Gamma$ limit.
The full evolution from the static ($a=0$ and $I=\pi\Gamma$) to the
deep shuttling regime ($a=5$, $I\approx 1$) is obtained
(cfr. also the current in Fig. \ref{FigNoise}).
The maximum of the distribution moves from approximately $\pi/10$ to 1.
Two features are particularly striking:
({\em i}) the peak becomes very sharp at the point that
a discontinuity  of the slope of $ \ln P$ appears at $\tilde n=1$
({\em ii}) it becomes asymmetric.
The fact that the peak is symmetric in
the static case is not surprising, the probability
of transferring more or less electrons than the
average should not be very different.
When $a$ becomes large we have instead shown that those
probabilities are controlled by two different
parameters, for $\tilde n<1$ by
$\alpha$ and for $\tilde n>1$ by $\tau = 2 \Gamma/a$.
The numerical results confirms this prediction.
The behavior around the maximum is well described
by the (nearly linear) form \refe{prob}.

\begin{figure}
\centerline{\psfig{file=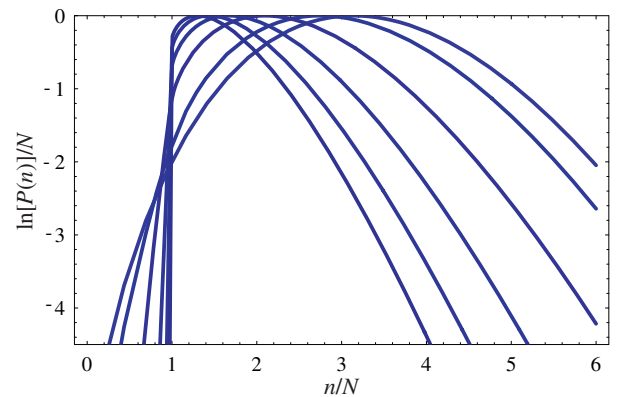,width=8cm}}
\caption{The same as Fig. \ref{FigFCS1} for $\Gamma=1$}
\label{FigFCS2}
\end{figure}

Figure \ref{FigFCS2} shows the case $\Gamma=1$.
In contrast with the previous case now for $a=0$
the maximum of the distribution if for
$\tilde n = 2 \pi$, larger than 1.
Shuttling will reduce the current to 1.
The main contribution to the transport comes from the
sequential hopping through the grain when both
$\Gamma_L$ and $\Gamma_R$ are non vanishing.
The oscillation reduces this region in favor of
regions where only one $\Gamma$ is non vanishing.
In this limit one electron per cycle is transferred.
Since this regime is attained when the contribution of the
region $x\approx 0$ becomes negligible, {\em i.e.}
when $\tau=2 \Gamma/a \rightarrow 0$, this means
that one needs huge oscillation amplitudes to reach
the truly shuttling regime of $\av{\tilde n}=1$.
For large $a$, but not yet in this limit, the
probability has the form shown in Fig. \ref{FigFCS2}.
We considered also this limit analytically after
\refE{largetau}.
Like in the previous cases a singularity develops at 1, but
in this case the probability remains monotonic at 1
(for not too large $a$).
The effect of the shuttling is thus mainly to enormously reduce
the probability that less than one electron is transferred, and
then to slightly shift the maximum in the distribution from
$\pi \Gamma>1$ towards 1.
This is due to the fact that due to the oscillations at least
one particle is always transferred  and the probability of
transferring more than one particle is reduced, since the time
spent by the shuttle in the central region $|x| \ll \lambda$ is
shorter.

\section{Conclusions}
\label{sec5}

In conclusion we have studied the full counting statistics
of charge transfer in a single
electron transition structure where the central grain can oscillate
at a given frequency.
The two relevant parameters are the oscillation amplitude divided by
the scale of the exponential dependence of the resistance ($a$), and
the probability of a tunnelling event during the time $1/\omega$
for the static structure ($\Gamma$).
We have obtained both numerical and analytical expressions for the FCS.
The results apply to both driven or self oscillating shuttles,
when the fluctuation of the amplitude of oscillation can be
neglected.
The probability of transferring $\tilde n$ electrons changes
qualitatively as a function of $a$ and $\Gamma$.
When $\Gamma>1$ the tunnelling events happening when
the shuttles passes through the region $x\approx 0$ are
always important and very large shuttling amplitudes are
necessary to have a well defined shuttling regime.

We also discussed in some details the first two moments
of the FCS: the current and the noise.
We found quantitative prediction for the reduction of the
Fano factor for large oscillation amplitudes.

The study of the FCS permits to understand more deeply
the dynamics of charge transfer.
In some cases we found that the effective elementary
charge becomes 1/2 the actual one, due to correlations.
This both in the static and in the dynamic regime.
In other limiting cases the statistics is in general polynomial,
taking into account the probability of different outcomes at each cycle.
Generalization of the theory to a larger number of available states in the
grain, or the inclusion of an asymmetric hopping probability is straightforward
and can be important to study more realistic systems.

\acknowledgements

I thank L.Y. Gorelik, Ya. M. Blanter, R.I. Shekhter, and Y.M. Galperin  for
useful discussion.
I also thank F.W.J. Hekking and M. Houzet for careful reading of the manuscript
and precious suggestions.
I acknowledge financial support from CNRS through contract ATIP-JC 2002.
This work was supported by the U.S. Department of Energy Office of Science via
contract No. W-31-109-ENG-38.

\end{document}